\documentclass[12pt]{article}
\usepackage[dvips]{graphicx}

\begin{document}

\title{Anomalous Effects of ``Guest'' Charges Immersed in 
Electrolyte: Exact 2D Results}

\author{L. {\v S}amaj$^1$}

\maketitle

\begin{abstract}
We study physical situations when one or two ``guest'' arbitrarily-charged
particles are immersed in the bulk of a classical electrolyte modelled by
a Coulomb gas of $\pm$ unit point-like charges, 
the whole system being in thermal equilibrium.
The models are treated as two-dimensional with logarithmic pairwise
interactions among charged constituents; the (dimensionless) inverse 
temperature $\beta$ is considered to be smaller than 2 in order to ensure 
the stability of the electrolyte against the collapse of positive-negative
pairs of charges.
Based on recent progress in the integrable (1+1)-dimensional 
sine-Gordon theory, exact formulas are derived for the chemical
potential of one guest charge and for the asymptotic large-distance
behavior of the effective interaction between two guest charges.
The exact results imply, under certain circumstances, anomalous
effects such as an effective attraction (repulsion) between like-charged 
(oppositely-charged) guest particles and the charge inversion in 
the electrolyte vicinity of a highly-charged guest particle.
The adequacy of the concept of renormalized charge is confirmed 
in the whole stability region of inverse temperatures 
and the related saturation phenomenon is revised.
\end{abstract}

\medskip


\vfill

\noindent $^1$ 
Institute of Physics, Slovak Academy of Sciences, D\'ubravsk\'a cesta 9, 
\newline 845 11 Bratislava, Slovak Republic; e-mail: fyzimaes@savba.sk

\newpage

\renewcommand{\theequation}{1.\arabic{equation}}
\setcounter{equation}{0}

\section{Introduction}
This paper deals with physical situations when one or two ``guest'' charges,
say arbitrarily charged colloidal particles with a hard core of radius 
$\sigma$, are immersed in a classical electrolyte modelled by an infinite 
Coulomb gas of positive/negative unit charges.
In order to obtain explicit results we consider the point-like limit
of the guest charges, i.e. $\sigma/\lambda \to 0$ where $\lambda$ is a
characteristic correlation length of electrolyte species; the obtained
results are not expected to be applicable to large-sized colloids.
If the charges of the guest particles are sufficiently large, 
anomalous counterintuitive phenomena emerge in the system \cite{Levin}.

One of such phenomena is the appearance, under some circumstances,
of an effective (i.e., mediated by the electrolyte) attraction 
between like-charged colloids.
While the traditional DLVO theory \cite{Derjaguin,Verwey} always
predicts an effective repulsion between two like-charged colloids
\cite{Crocker1,Chakrabarti}, experimental measurements \cite{Kepler,Crocker2}
and numerical simulations \cite{Jensen,Allahyarov} provide evidence for
attraction, especially within confined geometries (close to a dielectric
wall or between two glass plates) but also in the bulk 
of the electrolyte \cite{Ma}.
It was argued that the effective attraction of two like-charged colloids
in the presence of a single wall can arise also from a non-equilibrium
hydrodynamic effect \cite{Squires1,Squires2}.

Another interesting effect is the overcharging, or the charge inversion,
of a highly charged colloid \cite{Shklovskii}.
This effect occurs when the number of electrolyte counterions in the
vicinity of the colloidal surface becomes so high that the colloidal
bare charge is locally overcompensated.
The charge inversion has been observed experimentally by
electrophoresis \cite{Walker} and in simulations \cite{Messina1}.
Its theoretical explanation is based on Wigner-crystal theories 
\cite{Nguyen,Messina2}.

The latter effect is related to the concept of renormalized charge
\cite{Manning,Alexander,Lowen,Diehl,Trizac,Auboy}.
The true electric potential far from the colloid immersed in 
a weakly-coupled electrolyte is supposed to exhibit the Debye-H\"uckel form, 
but with a renormalized-charge prefactor which is different from 
the bare charge of the colloid.
An important feature, which occurs in the framework of the nonlinear
Poisson-Boltzmann equation, is that the renormalized charge saturates
monotonically at some finite value when the colloidal bare charge goes 
to infinity \cite{Trizac,Auboy}.
Monte-Carlo simulations of a salt-free colloidal cell model \cite{Groot} 
indicate the existence of a maximum in the plot 
of the renormalized charge versus the bare colloidal charge.
T\'ellez and Trizac \cite{Tellez} considered the possibility of
a more general phenomenon of {\em potential} saturation.

A theoretical elucidation of anomalous phenomena requires to go
beyond mean-field approximations by incorporating electrostatic
correlations among electrolyte particles.
Heuristic phenomenological approaches applied so far are based
on plausible, but not rigorously justified, arguments.
Some exactly solvable models are needed.
The best candidates are two-dimensional (2D) Coulomb systems
with logarithmic pairwise interactions among the charged constituents.
The 2D Coulomb gas of $\pm$ unit point-like charges is stable
against the collapse of positive-negative pairs of charges at
high enough temperatures, namely for $\beta<2$ where $\beta$ is
the (dimensionless) inverse temperature or coupling constant.
The collapse point $\beta=2$, at which the collapse starts to occur,
is equivalent to the free-fermion point of the Thirring representation 
of the 2D Coulomb gas \cite{Cornu1,Cornu2}; although the free energy 
and the particle density diverge, the truncated Ursell correlation functions 
are finite at this point.
In a recent work \cite{Samaj1}, we have solved exactly the 2D problem
of one colloid immersed in the Coulomb gas just at the free-fermion point.
An explicit form of the induced electric potential as a function of
the bare colloidal charge was derived at every point of the space.
Based on this exact result, the concept of renormalized charge was shown
to fail in this strong-coupling regime.
On the other hand, the anticipated phenomenon of the electric potential
saturation was confirmed at the free-fermion point.

Our present aim is to extend the exact treatment of the guest-charge(s)
problem to the whole Coulomb-gas stability region of inverse temperatures
$0\le\beta<2$.
A first important step towards this aim has already been done by solving
exactly the equilibrium statistical mechanics of the 2D Coulomb gas
in the stability regime (the bulk thermodynamics, special cases of the
surface thermodynamics and the large-distance behavior of the two-body
correlation functions) via an equivalence with the integrable 2D
Euclidean sine-Gordon theory; for a short review, see ref. \cite{Samaj2}.
As is shown in this paper, the problem of one (two) guest charge(s)
immersed in the Coulomb plasma is related to the evaluation of
one-point (two-point) expectation values of the exponential field
in the sine-Gordon theory.
Based on recent progress in the latter topic, we derive explicit
formulas for the chemical potential of one guest charge and for
the asymptotic large-distance behavior of the effective interaction
between two guest charges.
The exact results imply, under some circumstances, an effective
attraction (repulsion) between like-charged (oppositely-charged) 
guest particles and the charge inversion in the electrolyte
around a highly-charged guest particle.
The adequacy of the concept of renormalized charge is confirmed 
in the whole stability region $0\le\beta<2$.
The related saturation phenomenon is revised.

The paper is organized as follows.
Basic facts about the bulk properties of the 2D symmetric Coulomb gas 
are summarized in Section 2.
Section 3 deals with the problem of one guest charge in the electrolyte.
The effective interaction between two guest charges immersed in
the electrolyte is studied within a form-factor method for the
equivalent sine-Gordon model in Section 4.
Based on the exact results of Section 4, the concept of renormalized
charge and the related saturation phenomenon are tested in Section 5.
A brief recapitulation and some concluding remarks are given in Section 6.

\renewcommand{\theequation}{2.\arabic{equation}}
\setcounter{equation}{0}

\section{Basic facts about the 2D Coulomb gas}
We consider an infinite 2D plane $\Lambda$ of points ${\bf r}\in R^2$,
filled with a homogeneous medium of dielectric constant $=1$. 
The electrostatic potential $v$ at a point ${\bf r}$, induced by a unit 
charge at the origin ${\bf 0}$, is given by the 2D Poisson equation
\begin{equation} \label{2.1}
\Delta v({\bf r}) = - 2\pi \delta({\bf r})
\end{equation}  
The solution of this equation, subject to the boundary condition 
$\nabla v({\bf r})\to 0$ as $\vert {\bf r}\vert\to\infty$, reads
\begin{equation} \label{2.2}
v({\bf r}) = - \ln \left( \frac{\vert {\bf r}\vert}{r_0} \right),
\qquad {\bf r}\in R^2
\end{equation}
The free length constant $r_0$ will be set for simplicity to unity. 
This definition of the Coulomb potential in 2D maintains many
generic properties (e.g., sum rules \cite{Martin}) of ``real''
3D Coulomb fluids with the interaction potential
$v({\bf r}) = 1/\vert {\bf r}\vert$, ${\bf r}\in R^3$.

The symmetric Coulomb gas consists of two species of point-like 
particles with opposite unit charges $q_j\in \{ +1,-1 \}$;
to simplify the notation, the elementary charge $e$ is set to unity.
The bulk properties of the system in thermodynamic equilibrium are usually
treated within the grand canonical ensemble.
The ensemble is characterized by the (dimensionless) inverse temperature 
$\beta$, which plays the role of the coupling constant, and by the couple 
of equivalent particle fugacities $z_+ = z_- = z$.
Since the length scale $r_0$ in (\ref{2.2}) was set to unity, the true 
dimension of $z$ is $[{\rm length}]^{-2+(\beta/2)}$.
The grand partition function of the plasma is defined by
\begin{equation} \label{2.3}
\Xi(z) = \sum_{N_+,N_-=0}^{\infty} \frac{1}{N_+! N_-!}
\int_{\Lambda} \prod_{j=1}^N \left[ {\rm d}^2 r_j\, z_{q_j} \right] 
\exp \left\{ - \beta \sum_{j<k} q_j q_k v(\vert {\bf r}_j-{\bf r}_k \vert)
\right\}  
\end{equation}
where $N_+$ $(N_-)$ is the number of positively (negatively) 
charged particles and $N=N_++N_-$. 
The system is stable against the collapse of positive-negative pairs of unit 
point-like charges provided that the corresponding Boltzmann factor 
$r^{-\beta}$ is integrable at short distances in 2D, i.e. for $\beta<2$.
In what follows, we shall restrict ourselves to this stability region
of coupling constants.

To introduce the averaged many-particle densities, we denote by
$\langle \cdots \rangle_{\beta}$ the standard thermal average. 
At the one-particle level, one considers the number density of
particles of one sign
\begin{equation} \label{2.4}
n_{q}({\bf r})  =  \left\langle \sum_j \delta_{q,q_j} 
\delta({\bf r}-{\bf r}_j) \right\rangle_{\beta} , \qquad q = \pm 1
\end{equation}
Due to the charge symmetry and space homogeneity, $n_+ = n_- = n/2$ where 
$n$ is the total density of particles.
At the two-particle level, one considers the two-body number densities
\begin{equation} \label{2.5}
n_{qq'}({\bf r},{\bf r}') = \left\langle \sum_{j\ne k} 
\delta_{q,q_j} \delta({\bf r}-{\bf r}_j) \delta_{q',q_k} 
\delta({\bf r}'-{\bf r}_k) \right\rangle_{\beta} , 
\qquad q,q' = \pm 1
\end{equation}
which are translationally invariant, $n_{qq'}({\bf r},{\bf r}') 
\equiv n_{qq'}(\vert {\bf r}-{\bf r}'\vert)$.
The two-body densities decouple at asymptotically large distance onto 
the product of the corresponding one-body densities,
$\lim_{r\to\infty} n_{qq'}(r) = n_{q} n_{q'}$.
It is therefore natural to introduce the Ursell functions,
$U_{qq'}(r) = n_{qq'}(r) - n_{q} n_{q'}$, 
which go to 0 as $r\to\infty$.
It is also useful to consider the pair distribution functions
$g_{qq'}(r) = n_{qq'}(r)/(n_{q} n_{q'})$. 

The short-distance behavior of the two-body densities is dominated by 
the Boltzmann factor of the corresponding pair Coulomb potential 
\cite{Jancovici77,Hansen}.
In particular, the pair distribution functions behave like
\begin{eqnarray} 
g_{qq'}(r) & \sim & C_{qq'} 
r^{\beta qq'} \qquad \mbox{as $r\to 0$}
\label{2.6} \\
C_{qq'} & = & \exp\left[ \beta(\mu_{q}^{\rm ex}+
\mu_{q'}^{\rm ex} - \mu_{q+q'}^{\rm ex}) \right] \label{2.7}
\end{eqnarray}
Here, the excess (i.e. over ideal) chemical potential of the Coulomb-gas 
species is defined by
\begin{equation} \label{2.8}
\exp \left( \beta \mu_{q}^{\rm ex} \right) = 
\frac{z_{q}}{n_{q}} , \qquad q = \pm 1
\end{equation}
and $\mu_Q^{\rm ex}$ with arbitrarily-valued $Q$ represents an extended
definition of the excess chemical potential: 
$\mu_Q^{\rm ex}$ is the reversible work which has to be done in order 
to bring a particle of charge $Q$ (in units of the elementary charge $e$) 
from infinity into the bulk interior of the considered Coulomb gas.
In the case of oppositely-charged particles, formula (\ref{2.6}) reduces to
\begin{equation} \label{2.9}
g_{+-}(r) \sim \left( \frac{z_+}{n_+} \right) \left( \frac{z_-}{n_-} \right)
r^{-\beta} \qquad \mbox{as $r\to 0$}
\end{equation}

According to Eq. (\ref{2.1}), $-\Delta/(2\pi)$ is the inverse operator
of the Coulomb potential $v({\bf r})$.
The grand partition function of the 2D Coulomb gas (\ref{2.3}) can be thus 
turned via the Hubbard-Stratonovich transformation 
(see e.g. ref. \cite{Minnhagen}) into
\begin{equation} \label{2.10}
\Xi = \frac{\int {\cal D}\phi\, \exp[-S(z)]}{\int {\cal D}\phi\, \exp[-S(0)]}
\end{equation}
with
\begin{equation} \label{2.11}
S(z) = - \int_{\Lambda} {\rm d}^2 r
\left[ \frac{1}{16\pi} \phi \Delta \phi + 2 z \cos( b \phi ) \right] ,
\qquad b^2 =  \frac{\beta}{4} 
\end{equation}
being the 2D Euclidean action of the sine-Gordon model.
Here, $\phi({\bf r})$ is a real scalar field and $\int {\cal D}\phi$
denotes the functional integration over this field.
The fugacity $z$ is renormalized by the (diverging) self-energy term
$\exp[\beta v(0)/2]$, without changing the $z$-notation.
The one- and two-body densities of the charged particles in the plasma
are expressible as averages over the sine-Gordon action as follows
\begin{equation} \label{2.12}
n_{q} = z_{q} \langle {\rm e}^{{\rm i} q b \phi} \rangle , 
\qquad n_{qq'}(\vert {\bf r}-{\bf r}'\vert) = z_{q} z_{q'} \langle 
{\rm e}^{{\rm i}q b \phi({\bf r})} 
{\rm e}^{{\rm i}q' b \phi({\bf r}')} \rangle 
\end{equation}
With regard to Eq. (\ref{2.8}), it holds
\begin{equation} \label{2.13}
\exp(-\beta \mu_{q}^{\rm ex}) = \langle {\rm e}^{{\rm i} q b \phi} 
\rangle , \qquad q = \pm 1
\end{equation}
The short-distance behavior (\ref{2.9}) is equivalent to
\begin{equation} \label{2.14}
\langle {\rm e}^{{\rm i} b \phi({\bf r})} 
{\rm e}^{-{\rm i} b \phi({\bf r}')} \rangle \sim
\vert {\bf r}-{\bf r}' \vert^{-4 b^2} \qquad 
\mbox{as $\vert {\bf r}-{\bf r}' \vert \to 0$}
\end{equation}
Under this conformal normalization of the exponential field, the divergent 
self-energy factor (which renormalizes $z$) disappears from statistical 
relations calculated within the sine-Gordon representation.
The short-distance formula (\ref{2.14}) is the special case of a
more general relation
\begin{equation} \label{2.15}
\langle {\rm e}^{{\rm i} a \phi({\bf r})} 
{\rm e}^{{\rm i} a' \phi({\bf r}')} \rangle \sim
\langle {\rm e}^{{\rm i}(a+a')\phi} \rangle
\vert {\bf r}-{\bf r}' \vert^{4 a a'} \qquad 
\mbox{as $\vert {\bf r}-{\bf r}' \vert \to 0$}
\end{equation} 
valid in a restricted region of the parameters $a$ and $a'$ such that
the one-point average $\langle{\rm e}^{{\rm i}(a+a')\phi}\rangle$ be finite.

The sine-Gordon model (\ref{2.11}) is integrable \cite{Zamolodchikov1}.
Its particle spectrum consists of one soliton-antisoliton pair $(S,{\bar S})$
with equal masses $M$ and of $S-{\bar S}$ bound states (``breathers'')
$\{ B_j; j = 1,2,\ldots < 1/\xi \}$ whose number at a given $b^2$
depends on the inverse of the parameter
\begin{equation} \label{2.16}
\xi = \frac{b^2}{1-b^2} \qquad \left( = \frac{\beta}{4-\beta} \right)
\end{equation}
The mass of the $B_j$-breather is given by
\begin{equation} \label{2.17}
m_j = 2 M \sin \left( \frac{\pi \xi}{2} j \right)
\end{equation}
and this breather disappears from the spectrum just when $m_j = 2 M$.
The breathers exist only in a subinterval of the stability region 
$0\le b^2 < 1/2$ $(0\le \beta <2)$ of the point-like Coulomb gas.
The lightest $B_1$-breather, usually called the elementary one,
has the mass
\begin{equation} \label{2.18}
m_1 = 2 M \sin \left( \frac{\pi \xi}{2} \right)
\end{equation}
and disappears from the particle spectrum just at the free-fermion point 
$b^2 = 1/2$ $(\beta = 2)$.
The soliton-antisoliton pair is present in the spectrum up to the
Kosterlitz-Thouless transition point $b^2=1$ $(\beta=4)$ 
at which the sine-Gordon theory ceases to be massive.

The (dimensionless) specific grand potential $\omega$ of the 2D
Euclidean sine-Gordon model, defined by
\begin{equation} \label{2.19}
- \omega = \lim_{\vert \Lambda\vert \to \infty} 
\frac{1}{\vert\Lambda\vert} {\rm ln} \Xi
\end{equation}
was found in ref. \cite{Destri} by using the Thermodynamic Bethe ansatz:
\begin{equation} \label{2.20}
- \omega = \frac{m_1^2}{8 \sin(\pi\xi)}
\end{equation}
Under the conformal normalization of the exponential fields (\ref{2.14}), 
the relationship between the fugacity $z$ and the soliton/antisoliton 
mass $M$ was established in ref. \cite{Zamolodchikov2},
\begin{equation} \label{2.21}
z = \frac{\Gamma(b^2)}{\pi \Gamma(1-b^2)}
\left[ M \frac{\sqrt{\pi} \Gamma((1+\xi)/2)}{2 \Gamma(\xi/2)}
\right]^{2-2b^2}
\end{equation}
where $\Gamma$ stands for the Gamma function.
Note that the mass $M$ has dimension of an inverse length.
As a consequence of Eqs. (\ref{2.20}) and (\ref{2.21}), one has
\begin{equation} \label{2.22}
\langle {\rm e}^{{\rm i}b\phi} \rangle 
= \frac{1}{2} \frac{\partial (-\omega)}{\partial z}
= \frac{M^2}{8 z (1-b^2)} {\rm tg} \left( \frac{\pi \xi}{2} \right)
\end{equation}
Relations (\ref{2.21}) and (\ref{2.22}), together with the equality
\begin{equation} \label{2.23}
\frac{n}{2 z} = \langle {\rm e}^{{\rm i}b\phi} \rangle
\end{equation} 
determine explicitly the density-fugacity relationship and consequently 
the complete thermodynamics of the 2D Coulomb gas 
in the stability region \cite{Samaj3}. 

\renewcommand{\theequation}{3.\arabic{equation}}
\setcounter{equation}{0}

\section{One guest charge in the electrolyte}

Let us consider a point-like particle of charge $Q$ with $Q$ being 
an arbitrarily valued real number;
when $Q$ is interpreted as the valence it has to be an integer.
The charge is put into the bulk interior of the 2D electrolyte, 
say at the origin ${\bf 0}$.
The electrostatic potential induced by the guest charge at a point 
${\bf r}\in R^2$ is equal to $- Q {\rm ln}\vert {\bf r}\vert$.
Its effect on the constant species fugacities $z_{\pm}$ is the following:
$z_q\to z_q^{(1)}({\bf r}) = z \vert {\bf r} \vert^{\beta Q q}$.
The excess chemical potential of the guest particle is given by
\begin{equation} \label{3.1}
\exp\left( -\beta \mu_Q^{\rm ex} \right)
= \frac{\Xi[ z_q^{(1)}({\bf r}) ]}{\Xi(z)}
\end{equation}
where $\Xi[z_q^{(1)}({\bf r})]$ represents an obvious functional 
generalization of the definition (\ref{2.3}) of the grand partition 
function with position-dependent particle fugacities.
Performing the Hubbard-Stratonovich transformation, the sine-Gordon 
representation of $\Xi[z_q^{(1)}({\bf r})]$ takes the standard form of 
Eq. (\ref{2.10}) with the action
\begin{equation} \label{3.2}
S^{(1)}(z) = - \int_{\Lambda} {\rm d}^2 r
\left( \frac{1}{16\pi} \phi \Delta \phi +
z \vert {\bf r} \vert^{4 Q b^2} {\rm e}^{{\rm i}b\phi} +
z \vert {\bf r} \vert^{-4 Q b^2} {\rm e}^{-{\rm i}b\phi} \right)
\end{equation}
We first shift the scalar field $\phi({\bf r}) \to \phi'({\bf r})
= \phi({\bf r}) - {\rm i} 4 Q b\, {\rm ln}\vert {\bf r} \vert$.
Using subsequently the Poisson equation 
$\Delta {\rm ln}\vert {\bf r} \vert = 2 \pi \delta({\bf r})$
and integrating by parts with vanishing boundary contributions at
$\vert {\bf r} \vert \to \infty$, one gets
\begin{equation} \label{3.3}
\exp\left( - \beta \mu_Q^{\rm ex} \right) =
\langle {\rm e}^{{\rm i}Q b \phi} \rangle
\end{equation} 
where the average is taken with the usual sine-Gordon action $S(z)$
given by Eq. (\ref{2.11}).
When $Q=\pm 1$, one recovers the previous result (\ref{2.13})
derived for the plasma constituents.
Note the obvious symmetry $\mu_Q^{\rm ex} = \mu_{-Q}^{\rm ex}$.

An exact formula for the expectation value of the exponential field
$\langle {\rm e}^{{\rm i}a\phi} \rangle$, where the sine-Gordon parameter
$b^2$ lies inside the stability region $0\le b^2<1/2$
and $a$ is a free real parameter, was conjectured by Lukyanov
and Zamolodchikov in ref. \cite{Lukyanov1}. 
In terms of our notation $a = Q b$ [see Eq. (\ref{3.3})], their formula reads
\begin{equation} \label{3.4}
\langle {\rm e}^{{\rm i}Q b \phi} \rangle =
\left[ \frac{\pi z \Gamma(1-b^2)}{\Gamma(b^2)} \right]^{\frac{Q^2 b^2}{1-b^2}} 
\exp\left[ I_b(Q) \right] , \qquad \vert Q \vert < \frac{1}{2 b^2}
\end{equation}
with
\begin{equation} \label{3.5}
I_b(Q) = \int_0^{\infty} \frac{{\rm d}t}{t} \left[
\frac{{\rm sinh}^2(2 Q b^2 t)}{2\, {\rm sinh}(b^2 t)\, {\rm sinh} t\,
{\rm cosh}[(1-b^2)t]} - 2 Q^2 b^2 {\rm e}^{-2 t} \right]
\end{equation}
The formula was ``guessed'' on the base of three exactly solvable
cases of the sine-Gordon theory: the semi-classical limit $b^2\to 0$,
the free-fermion point $b^2=1/2$ \cite{Bernard} and the special value
of $a=b$, see Eq. (\ref{2.22}).
The validity of the formula was supported later by a ``reflection''
relationship with the imaginary Liouville theory \cite{Fateev},
a numerical study of the sine-Gordon model in finite volume
\cite{Bajnok} and a variational perturbation theory \cite{Lu}.
Other checks, provided by the Coulomb-gas representation, are presented
in the next two paragraphs.

The integral (\ref{3.5}) is finite provided that $\vert Q\vert < 1/(2b^2)$;
at $\vert Q\vert = 1/(2b^2)$, the integrated function behaves like $1/t$
for $t\to\infty$ what causes the logarithmic divergence of the integral.
In the Coulomb-gas picture, the interaction Boltzmann factor of the
$Q$-charge with an opposite unit plasma charge (counterion) at distance 
$r$, $r^{-\beta\vert Q\vert}$, is integrable at small $r$ in 2D 
if and only if $\beta \vert Q\vert < 2$.
The stability region for $\mu_Q^{\rm ex}$ therefore is expected to be
$\vert Q\vert < 2/\beta$; there is a collapse at $\vert Q\vert = 2/\beta$ 
characterized by $\mu_Q^{\rm ex}\to -\infty$.
With regard to the relationship $\beta = 4 b^2$ we conclude that the pair of
Eqs. (\ref{3.4}) and (\ref{3.5}) passes the guest-charge collapse test.

The way in which $\langle {\rm e}^{{\rm i}Q b \phi} \rangle$ diverges 
as $\vert Q\vert$ approaches the collapse value $1/(2b^2)$ is another 
check provided by the Coulomb-gas representation.
To show this fact, let us attach a hard core of radius $\sigma$ around
the guest $Q$-charge.
The effect of the $Q$-charge on the species fugacities is the following:
$z_q\to z_q({\bf r}) = z_q \vert {\bf r} \vert^{\beta Q q}\, 
\theta(\vert {\bf r}\vert - \sigma)$, where $\theta$ is 
the Heaviside step function.
The procedure analogous to that outlined between Eqs. (\ref{3.1})
and (\ref{3.3}) leads to
\begin{equation} \label{3.6}
\exp\left[ - \beta \mu_Q^{\rm ex}(\sigma) \right] =
\langle {\rm e}^{{\rm i}Q b \phi({\bf 0})} \rangle_{\sigma}
\end{equation}
where the average is taken with the action
\begin{equation} \label{3.7}
S_{\sigma}(z) = S(z) + z \int_{r<\sigma} {\rm d}^2 r \left[ 
{\rm e}^{{\rm i} b \phi({\bf r})} + {\rm e}^{-{\rm i} b \phi({\bf r})} \right]
\end{equation}
We expand $\langle {\rm e}^{{\rm i}Q b \phi({\bf 0})} \rangle_{\sigma}$
in the lowest $\sigma$-order around the sine-Gordon action $S(z)$:
\begin{equation} \label{3.8}
\langle {\rm e}^{{\rm i}Q b \phi({\bf 0})} \rangle_{\sigma} =
\langle {\rm e}^{{\rm i}Q b \phi} \rangle - z \int_{r<\sigma} 
{\rm d}^2 r \left[ \langle {\rm e}^{{\rm i}Q b \phi({\bf 0})} 
{\rm e}^{{\rm i}b\phi({\bf r})} \rangle +
\langle {\rm e}^{{\rm i}Q b \phi({\bf 0})} 
{\rm e}^{-{\rm i}b\phi({\bf r})} \rangle \right] + \cdots
\end{equation}
Due to the symmetry $\mu_Q^{\rm ex}(\sigma) 
= \mu_{-Q}^{\rm ex}(\sigma)$, it is sufficient to consider the case $Q>0$.
Applying in Eq. (\ref{3.8}) the short-distance formula (\ref{2.15}), 
the leading $\sigma$-correction is obtained in the form
\begin{equation} \label{3.9}
\langle {\rm e}^{{\rm i}Q b \phi({\bf 0})} \rangle_{\sigma} \sim
\langle {\rm e}^{{\rm i}Q b \phi} \rangle -
\pi z \langle {\rm e}^{{\rm i}(Q-1)b\phi} \rangle
\frac{\sigma^{2-4 Q b^2}}{1-2 Q b^2} \qquad \mbox{as $\sigma\to 0$}
\end{equation}
Close to the collapse value of the guest charge, i.e. when
$Q = 1/(2b^2) - \epsilon$ with $\epsilon\to 0^+$, one expands
\begin{equation} \label{3.10}
\frac{\sigma^{2-4 Q b^2}}{1- 2 Q b^2} 
= \frac{\sigma^{4 b^2 \epsilon}}{2 b^2 \epsilon}
= \frac{1}{2 b^2 \epsilon} + 2\, {\rm ln} \sigma + O(\epsilon)
\end{equation}
Since the regularized $\langle {\rm e}^{{\rm i}Q b \phi({\bf 0})} 
\rangle_{\sigma}$ has to be finite at the collapse $Q = 1/(2b^2)$, 
it follows from Eqs. (\ref{3.9}) and (\ref{3.10}) that the singular behavior
\begin{equation} \label{3.11}
\lim_{\epsilon\to 0^+} \left\langle 
\exp\left[ {\rm i}\left( \frac{1}{2b}
-\epsilon b \right) \phi \right] \right\rangle
\sim \frac{\pi z}{2 b^2 \epsilon} 
\left\langle \exp\left[ {\rm i}\left( \frac{1}{2b} 
- b \right) \phi \right] \right\rangle
\end{equation}
must take place. 
It is shown in the Appendix that the singular behavior (\ref{3.11}) is 
indeed reproduced by the conjectured Eqs. (\ref{3.4}) and (\ref{3.5}).

\renewcommand{\theequation}{4.\arabic{equation}}
\setcounter{equation}{0}

\section{Two guest charges in the electrolyte}
Let us put two point-like particles into the bulk of the Coulomb plasma,
the one with the charge $Q_1$ at the point ${\bf r}_1$ and the other with
the charge $Q_2$ at the point ${\bf r}_2$.
The electrostatic potential induced by these two charges at a point
${\bf r}\in R^2$ is equal to $-Q_1 {\rm ln}\vert {\bf r}-{\bf r}_1\vert 
-Q_2 {\rm ln}\vert {\bf r}-{\bf r}_2\vert$. 
The constant species fugacities $z_{\pm}$ are thus modified as follows:
$z_q \to z_q^{(2)}({\bf r}) = z \vert {\bf r}-{\bf r}_1\vert^{\beta Q_1 q} 
\vert {\bf r}-{\bf r}_2\vert^{\beta Q_2 q}$.
The excess chemical potential of the guest 1,2-charges as a whole
is given by
\begin{equation} \label{4.1}
\exp \left[ -\beta \mu_{Q_1 Q_2}^{\rm ex}({\bf r}_1,{\bf r}_2) \right]
= \vert {\bf r}_1-{\bf r}_2\vert^{\beta Q_1 Q_2}
\frac{\Xi[z_q^{(2)}({\bf r})]}{\Xi(z)} 
\end{equation}
Under the Hubbard-Stratonovich transformation, the sine-Gordon representation
of $\Xi[z_q^{(2)}({\bf r})]$ takes the standard functional form 
of Eq. (\ref{2.10}) with the action
\begin{eqnarray}
S^{(2)}(z) & = & - \int_{\Lambda} {\rm d}^2 r \Bigg(
\frac{1}{16\pi} \phi \Delta \phi +
z \vert {\bf r}-{\bf r}_1\vert^{4 Q_1 b^2} 
\vert {\bf r}-{\bf r}_2\vert^{4 Q_2 b^2} {\rm e}^{{\rm i}b\phi} \nonumber \\
& & + z \vert {\bf r}-{\bf r}_1\vert^{- 4 Q_1 b^2} 
\vert {\bf r}-{\bf r}_2\vert^{- 4 Q_2 b^2} {\rm e}^{- {\rm i}b\phi} \Bigg)
\label{4.2}
\end{eqnarray}
Shifting the scalar field $\phi({\bf r}) \to \phi'({\bf r}) = \phi({\bf r})
- {\rm i} 4 Q_1 b {\rm ln}\vert {\bf r}-{\bf r}_1 \vert
- {\rm i} 4 Q_2 b {\rm ln}\vert {\bf r}-{\bf r}_2 \vert$,
applying the Poisson equations $\Delta {\rm ln} \vert {\bf r}-{\bf r}_j \vert
= 2 \pi \delta({\bf r}-{\bf r}_j)$ $(j=1,2)$ and integrating by parts,
one arrives at
\begin{equation} \label{4.3}
\exp \left[ -\beta \mu_{Q_1 Q_2}^{\rm ex}(\vert {\bf r}_1
-{\bf r}_2)\vert \right] = \langle {\rm e}^{{\rm i} Q_1 b \phi({\bf r}_1)} 
{\rm e}^{{\rm i} Q_2 b \phi({\bf r}_2)} \rangle
\end{equation}
where the average is taken with the usual sine-Gordon action (\ref{2.11}).
The effective interaction energy between the guest 1,2-charges is defined by 
\begin{equation} \label{4.4}
E_{Q_1 Q_2}(\vert {\bf r}_1-{\bf r}_2 \vert) =
\mu_{Q_1 Q_2}^{\rm ex}(\vert {\bf r}_1-{\bf r}_2 \vert)
- \mu_{Q_1}^{\rm ex} - \mu_{Q_2}^{\rm ex} 
\end{equation}
With respect to the relation (\ref{3.3}), it holds
\begin{equation} \label{4.5}
\exp\left[ - \beta E_{Q_1 Q_2}(\vert {\bf r}_1-{\bf r}_2 \vert)\right] 
= \frac{\langle {\rm e}^{{\rm i} Q_1 b \phi({\bf r}_1)}
{\rm e}^{{\rm i} Q_2 b \phi({\bf r}_2)} \rangle}{
\langle {\rm e}^{{\rm i} Q_1 b \phi}\rangle 
\langle {\rm e}^{{\rm i} Q_2 b \phi}\rangle}
\end{equation}
At asymptotically large distance $\vert {\bf r}_1-{\bf r}_2\vert \to \infty$,
the two-point correlator $\langle {\rm e}^{{\rm i} Q_1 b \phi({\bf r}_1)}
{\rm e}^{{\rm i} Q_2 b \phi({\bf r}_2)} \rangle$ decouples onto the product
$\langle {\rm e}^{{\rm i} Q_1 b \phi}\rangle 
\langle {\rm e}^{{\rm i} Q_2 b \phi}\rangle$ and so the interaction energy 
goes to zero as it should be.
From Eq. (\ref{4.5}) one then gets
\begin{equation} \label{4.6}
- \beta E_{Q_1 Q_2}(\vert {\bf r}_1-{\bf r}_2 \vert)
\sim \frac{\langle {\rm e}^{{\rm i} Q_1 b \phi({\bf r}_1)}
{\rm e}^{{\rm i} Q_2 b \phi({\bf r}_2)} \rangle}{
\langle {\rm e}^{{\rm i} Q_1 b \phi}\rangle 
\langle {\rm e}^{{\rm i} Q_2 b \phi}\rangle} -1 ,
\qquad \vert {\bf r}_1-{\bf r}_2 \vert \to \infty
\end{equation} 
This means that the asymptotic large-distance behavior of the effective 
interaction energy between the guest particles is related to the 
large-distance behavior of the corresponding two-point correlation function 
of exponential fields associated with the 2D sine-Gordon theory.
 
For the 2D Euclidean sine-Gordon model, like for any integrable 2D theory,
the two-point correlation function of local operators ${\cal O}_a$ 
($a$ is a free parameter) can be formally expressed as an infinite 
convergent series over multi-particle intermediate states \cite{Smirnov},
\begin{eqnarray} 
\langle {\cal O}_a({\bf r}) {\cal O}_{a'}({\bf r}')\rangle 
& = & \sum_{N=0}^{\infty}
\frac{1}{N!} \sum_{\epsilon_1,\ldots,\epsilon_N} \int_{-\infty}^{\infty} 
\frac{{\rm d}\theta_1 \cdots {\rm d}\theta_N}{(2\pi)^N}
F_a(\theta_1,\ldots,\theta_N)_{\epsilon_1\cdots \epsilon_N} 
\phantom{space}\nonumber \\
& & {^{\epsilon_N\cdots \epsilon_1}F}_{a'}(\theta_N,\ldots,\theta_1)
\exp\left( - \vert {\bf r}-{\bf r}' \vert \sum_{j=1}^N m_{\epsilon_j}
{\rm cosh} \theta_j \right) \label{4.7}
\end{eqnarray}
where $\epsilon$ indexes the particles [say $\epsilon = + (-)$ for
a soliton (antisoliton) and $\epsilon = j$ for a $B_j$-breather] and
the rapidity $\theta\in (-\infty,\infty)$ parametrizes the energy
and the momentum of the corresponding particle.
The form factors
\begin{eqnarray}
F_a(\theta_1,\ldots,\theta_N)_{\epsilon_1\cdots \epsilon_N} 
& = & \langle 0 \vert {\cal O}_a({\bf 0}) \vert Z_{\epsilon_1}(\theta_1),
\ldots, Z_{\epsilon_N}(\theta_N) \rangle  \label{4.8} \\
{^{\epsilon_N\cdots \epsilon_1}F}_{a'}(\theta_N,\ldots,\theta_1)
& = & \langle Z_{\epsilon_N}(\theta_N),\ldots,Z_{\epsilon_1}(\theta_1)\vert
{\cal O}_{a'}({\bf 0}) \vert 0 \rangle \label{4.9}
\end{eqnarray}
are the matrix elements of the operator at the origin, between an
$N$-particle in-state (being a linear superposition of free one-particle
states $\vert Z_{\epsilon}(\theta)\rangle$) and the vacuum.
The first $N=0$ term of the series expansion (\ref{4.7}) corresponds
to the decoupling $\langle {\cal O}_a \rangle \langle {\cal O}_{a'} \rangle$.

The form-factor representation (\ref{4.7}) is particularly useful 
in the limit $\vert {\bf r}-{\bf r}' \vert \to \infty$ where the dominant
contribution to the truncated correlation function
$\langle {\cal O}_a({\bf r}) {\cal O}_{a'}({\bf r}')\rangle 
- \langle {\cal O}_a \rangle \langle {\cal O}_{a'} \rangle$
comes from a multi-particle state with the minimum value of the total
particle mass $\sum_{j=1}^N m_{\epsilon_j}$, at the point of vanishing
rapidities.
As was already mentioned, the lightest particle in the stability region 
$0\le b^2 < 1/2$ is the elementary breather $B_1$.
For this particle, the one-particle form factors $F_a(\theta)_1$
and ${^1F}_{a'}(\theta) = F_{a'}(\theta)_1$ of the exponential operator
${\cal O}_a({\bf r}) = \exp\left[ {\rm i} a \phi({\bf r}) \right]$
were calculated in refs. \cite{Lukyanov2,Lukyanov3}:
\begin{equation} \label{4.10}
F_a(\theta)_1 \equiv \langle 0 \vert {\rm e}^{{\rm i}a\phi} \vert B_1(\theta)
\rangle = - {\rm i} \langle {\rm e}^{{\rm i}a\phi} \rangle
\sqrt{\pi\lambda} \frac{\sin(\pi\xi a/b)}{\sin(\pi \xi)}
\end{equation}
where
\begin{equation} \label{4.11}
\lambda = \frac{4}{\pi} \sin(\pi \xi) \cos\left( \frac{\pi \xi}{2} \right)
\exp\left\{ - \int_0^{\pi\xi} \frac{{\rm d}t}{\pi} \frac{t}{\sin t} \right\}
\end{equation}
and $\xi$ is defined in Eq. (\ref{2.16}).
Since the form factor (\ref{4.10}) does not depend on the rapidity, 
the integration over $\theta$ in (\ref{4.7}) can be done explicitly 
by using the relation
\begin{equation} \label{4.12}
\int_{-\infty}^{\infty} \frac{{\rm d}\theta}{2}
{\rm e}^{- r m_1 {\rm cosh}\theta} = K_0(m_1 r) \sim
\left( \frac{\pi}{2 m_1 r} \right)^{1/2} \exp( -m_1 r)
\qquad \mbox{as $r\to\infty$}
\end{equation}
where $K_0$ is the modified Bessel function of second kind \cite{Gradshteyn}.
On the base of the asymptotic equivalence (\ref{4.6}), the large-distance
behavior of the effective interaction energy is finally found in the form
\begin{equation} \label{4.13}
\beta E_{Q_1 Q_2}(r) \sim [ Q_1 ] [ Q_2 ] \lambda 
\left( \frac{\pi}{2 m_1 r} \right)^{1/2} \exp( -m_1 r) ,
\qquad \mbox{$r\to\infty$}
\end{equation}
Here, the symbol $[ Q ]$ stands for the ratio
\begin{equation} \label{4.14}
[ Q ] = \frac{\sin\left( \pi \beta Q/(4-\beta) \right)}{\sin\left( 
\pi \beta/(4-\beta) \right)}
\end{equation}
Using the thermodynamic formulas derived at the end of Section 2,
the mass $m_1$ (which plays the role of the inverse charge-charge 
correlation length of the Coulomb-gas particles) is expressible as
\begin{equation} \label{4.15}
m_1 = \kappa \left[ \frac{\sin\left(\pi\beta/(4-\beta)\right)}{
\pi\beta/(4-\beta)} \right]^{1/2}
\end{equation}
where $\kappa=\sqrt{2\pi\beta n}$ denotes the inverse Debye length.
The $\beta$-dependence of the parameter $\lambda$, defined by 
Eq. (\ref{4.11}), reads
\begin{equation} \label{4.16}
\lambda = \frac{4}{\pi} \sin\left( \frac{\pi\beta}{4-\beta} \right)
\cos\left( \frac{\pi\beta}{2(4-\beta)} \right)
\exp\left\{ - \int_0^{\frac{\pi\beta}{4-\beta}} \frac{{\rm d}t}{\pi}
\frac{t}{\sin t} \right\}  
\end{equation}
An interesting feature of the result (\ref{4.13}) is that the effective
interaction energy factorizes into the product of separate charge 
contributions $[Q]$ from each of the guest particles. 

We would like to emphasize that the asymptotic formula (\ref{4.13}) 
was derived for the guest particles of point-like nature.
Its rigorous validity is therefore restricted to such guest charges
which do not collapse with an opposite unit counterion from the electrolyte, 
i.e. to the values $\vert Q_1\vert,\vert Q_2\vert < 2/\beta$.
Since the function $[Q]$ is analytic at every real $Q$, 
the stability border $\vert Q\vert =2/\beta$ does not represent 
an exceptional point at which a singularity emerges 
[like it was in the case of the excess chemical potential
determined by Eqs. (\ref{3.3})-(\ref{3.5})].
The explanation of this important fact follows from the definition 
(\ref{4.4}) of the effective interaction energy: if one of the guest
charges passes or is beyond its collapse value, say
$\vert Q_1\vert \ge 2/\beta$, both excess chemical potentials 
$\mu_{Q_1Q_2}^{\rm ex}(\vert {\bf r}_1-{\bf r}_2\vert)$ and 
$\mu_{Q_1}^{\rm ex}$ tend to $-\infty$ in such a way that their
difference is expected to keep a finite value.
We therefore suggest that the formula (\ref{4.13}) remains valid for 
arbitrary real values of $Q_1$ and $Q_2$, and corresponds to the
limit of a small hard-core radius $\sigma$ (such that $m_1 \sigma <<1$) 
around the guest charges.
In what follows, we shall refer to this conjecture as 
``the regularization hypothesis''.

The interaction energy $E_{Q_1 Q_2}(r)$ is repulsive (attractive) at 
asymptotically large distance $r$ when it goes to zero from above (below).
When the amplitudes of the guest charges are the same, 
i.e. $\vert Q_1\vert = \vert Q_2\vert$, the interaction energy (\ref{4.13})
exhibits the vacuum-type behavior: it is repulsive for $Q_1 = Q_2$
and, since $[-Q]=-[Q]$, attractive for $Q_1 = -Q_2$.
The situation is more complex when the amplitudes of the guest charges 
differ from one another.
Let us analyze the plot of $[Q]$ as the function of (say positive) 
bare charge $Q$ for a fixed value of $\beta$ from the stability
range $\langle 0,2)$. 
In the interval $Q\in\langle 0,(2/\beta)-(1/2) \rangle$, $[Q]$ increases
monotonically from 0 to its maximum at the end-point of this interval.
In the subsequent interval $Q\in\langle (2/\beta)-(1/2),(4/\beta)-1 \rangle$,
which contains the stability border $Q=2/\beta$, $[Q]$ is the decreasing
function of $Q$ but still keeps the positive sign of $Q$.
Such behavior means physically that by increasing the bare charge of one 
of the guest particles the effective interaction energy weakens 
which is a counterintuitive phenomenon.
The function $[Q]$ changes the sign when $Q$ passes the value 
$(4/\beta)-1$; under the assumption of validity of the
regularization hypothesis, this indicates an effective change of
the sign of the bare charge $Q$.  
Appropriate combinations of the $Q_1,Q_2$-charges in the factorized
relation (\ref{4.13}) can therefore lead to an effective attraction
(repulsion) between like-charged (oppositely-charged) guest particles.
The described scenario repeats itself when increasing $Q$ due to 
the periodicity relation $[Q] = [Q+(8/\beta)-2]$. 

\begin{figure}[h]
\begin{center}
\includegraphics{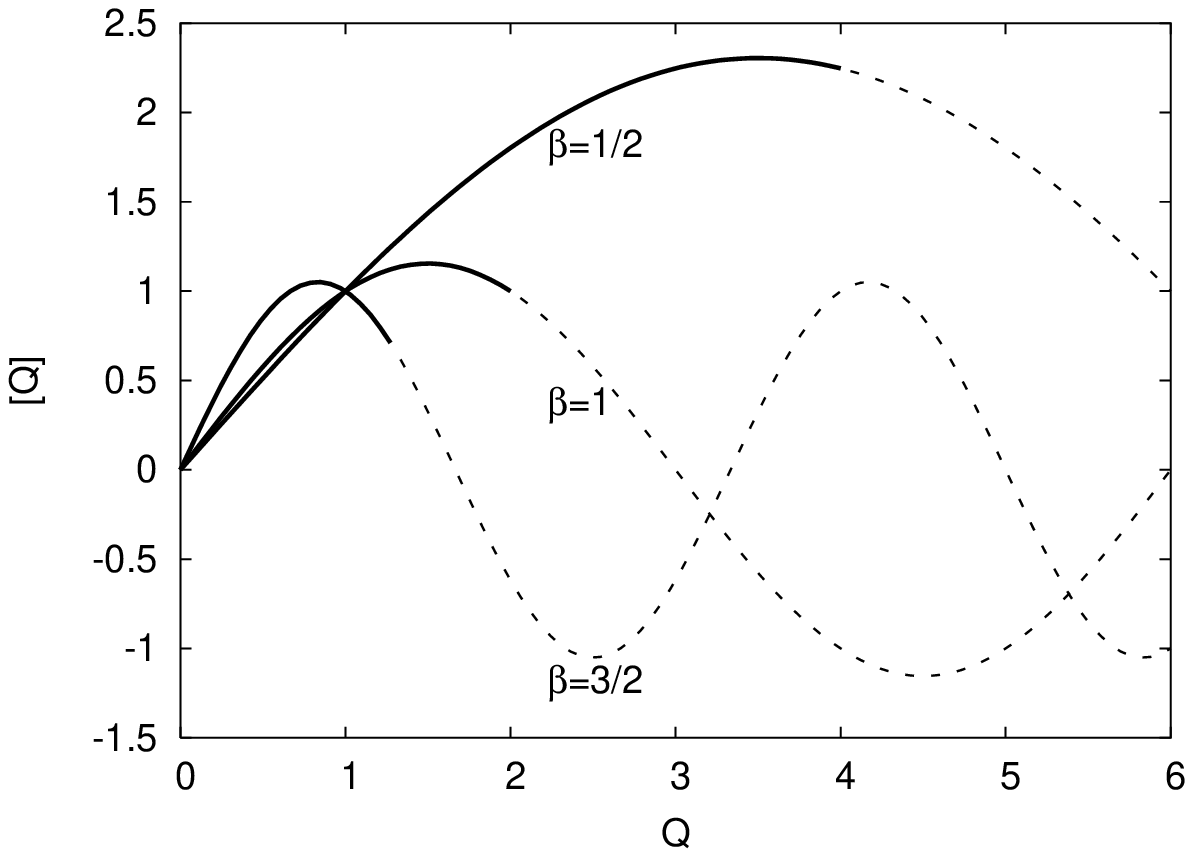}
\mbox{Fig. 1. The function $[Q]$ versus the bare charge $Q$ for
various values of $\beta$.}
\end{center}
\end{figure}

The above discussed plot of $[Q]$ versus the (positive) bare charge $Q$
is presented graphically for various values of the electrolyte coupling 
constant $\beta$ in Fig. 1.
The solid-line parts of the plots correspond to the interval $0<Q<2/\beta$ 
for which the formula (\ref{4.13}) is valid rigorously, the dashed-line parts 
of the plots correspond to $Q\ge 2/\beta$ when the formula (\ref{4.13}) 
is applicable under the assumption of the regularization hypothesis.

\renewcommand{\theequation}{5.\arabic{equation}}
\setcounter{equation}{0}

\section{Renormalized charge}
Let us put one point-like particle of charge $Q$ at the origin ${\bf 0}$
and look for the evoked density profiles $n_q({\bf r})$ of the
electrolyte species $q=\pm 1$.
One can formally follow the procedure for two guest particles
(one of which has the charge $q=\pm 1$ of the electrolyte species) 
outlined at the beginning of Section 4, to obtain
\begin{equation} \label{5.1}
n_q({\bf r}) = z \frac{\langle {\rm e}^{{\rm i}Q b \phi({\bf 0})}
{\rm e}^{{\rm i}q b \phi({\bf r})} \rangle}{\langle
{\rm e}^{{\rm i}Q b \phi} \rangle} , \qquad q = \pm 1
\end{equation}
Note that in the special case $Q=0$ the spatially-homogeneous relation
(\ref{2.23}) is reproduced.
At asymptotically large distance $r$ from the guest $Q$-charge,
the straightforward application of the form-factor method explained
in Section 4 provides the result
\begin{equation} \label{5.2}
n_q(r) \sim n_q \left\{ 1 - [q] [Q] \lambda
\left( \frac{\pi}{2 m_1 r} \right)^{1/2} \exp ( - m_1 r ) \right\} ,
\qquad r \to \infty
\end{equation}
where the symbols $[q]$, $[Q]$ are defined by Eq. (\ref{4.14}) and
the parameters $m_1$, $\lambda$ by the respective relations
(\ref{4.15}) and (\ref{4.16}).

The charge density $\rho$ of electrolyte unit charges is defined as
$\rho({\bf r}) = n_+({\bf r}) - n_-({\bf r})$.
Since $[q] = q$ for $q=\pm 1$, one finds that
\begin{equation} \label{5.3}
\rho(r) \sim - n [Q] \lambda \left( \frac{\pi}{2 m_1 r} \right)^{1/2} 
\exp ( - m_1 r ) , \qquad r \to \infty
\end{equation} 

The average electrostatic potential $\psi$ induced by the guest $Q$-charge
is related to the charge-density profile through the 2D Poisson equation,
\begin{equation} \label{5.4}
\Delta \psi({\bf r}) = - 2 \pi \rho({\bf r})
\end{equation}
Inserting here the asymptotic formula (\ref{5.3}) and considering 
the circularly symmetric Laplacian 
$\Delta = r^{-1} {\rm d}_r ( r {\rm d}_r )$, the dimensionless
electric potential $\beta \psi$ can be shown to behave at large distance
$r$ from the $Q$-charge as follows
\begin{equation} \label{5.5}
\beta \psi(r) \sim \left( \frac{\kappa}{m_1} \right)^2 [Q] \lambda
\left( \frac{\pi}{2 m_1 r} \right)^{1/2} 
\exp ( - m_1 r ) , \qquad r \to \infty
\end{equation}
In the Debye-H\"uckel limit $\beta\to 0$ and for finite $Q$,
it holds $[Q]\sim Q$, $m_1\sim \kappa$ and $\lambda\sim \beta$.
Eq. (\ref{5.5}) thus reduces to the well-known result 
(see e.g. refs. \cite{Levin,Samaj1})
\begin{equation} \label{5.6}
\beta \psi_{\rm DH}(r) \sim \beta Q 
\left( \frac{\pi}{2 \kappa r} \right)^{1/2} 
\exp ( - \kappa r ) , \qquad r \to \infty
\end{equation}  
The asymptotic behaviors (\ref{5.5}) and (\ref{5.6}), considered
respectively in terms of the dimensionless combinations
$m_1 r$ and $\kappa r$, exhibit the same type of the fall-off.
This fact confirms the adequacy of the concept of renormalized charge
\cite{Manning}--\cite{Tellez} in the stability weak-coupling
regime of the Coulomb gas.
Eq. (\ref{5.5}) is consistent with Eq. (\ref{5.6}) provided that
one introduces the renormalized charge $Q_{\rm ren}$ as follows
\begin{equation} \label{5.7}
\beta Q_{\rm ren} = \left( \frac{\kappa}{m_1} \right)^2 [Q] \lambda
\end{equation} 
This formula can be simplified to
\begin{equation} \label{5.8}
Q_{\rm ren}(\beta,Q) = A(\beta) 
\sin\left( \frac{\pi\beta Q}{4-\beta}\right)
\end{equation}
where the positive amplitude $A(\beta)$ is given by
\begin{equation} \label{5.9}
\frac{1}{A(\beta)} = \frac{1}{2} (4-\beta)
\sin\left( \frac{\pi\beta}{2(4-\beta)} \right)
\exp\left\{ \int_0^{\frac{\pi\beta}{4-\beta}} \frac{{\rm d}t}{\pi}
\frac{t}{\sin t} \right\}
\end{equation} 

As was mentioned above, because of the point-like nature of the guest 
$Q$-charge the rigorous validity of the result (\ref{5.8}) is restricted 
to $\vert Q\vert < 2/\beta$.
Changing $Q$ from 0 towards positive real values, the renormalized charge
(\ref{5.8}) increases monotonically up to its maximum $A(\beta)$ at the
point $Q=(2/\beta)-(1/2)$.
Increasing then $Q$ from $(2/\beta)-(1/2)$ up to the stability border
$2/\beta$, $Q_{\rm ren}$ paradoxically decreases while still keeping 
the positive sign of the bare charge $Q$.
This scenario resembles the one observed in the Groot's Monte-Carlo
simulations of the salt-free (only counterions are present) colloidal 
cell model \cite{Groot}.

Under the assumption of validity of the regularization hypothesis,
one can further increase the value of $Q$ beyond $2/\beta$ in 
the relation (\ref{5.8}).
$Q_{\rm ren}$ changes its positive sign to the negative one at $Q=(4/\beta)-1$.
This change of the sign is closely related to the effect of
charge inversion \cite{Shklovskii}--\cite{Messina2}:
since the total screening cloud of electrolyte particles must compensate
exactly the bare charge $Q$ of the guest particle (the electroneutrality
sum rule), the fact that the charge density (\ref{5.3}) goes to 0
at asymptotically large distance $r$ from above is the evidence of 
the charge inversion starting at some distance from the guest $Q$-charge.

The renormalized charge $Q_{\rm ren}$ is a periodic function of $Q$.
This is why going with $Q\to \infty$ does not imply the saturation of 
$Q_{\rm ren}$ at some finite value.
Instead, $Q_{\rm ren}$ oscillates between the two finite $\pm A(\beta)$
extremes.
With regard to Eq. (\ref{5.5}), the induced electrostatic potential
exhibits the same type of the oscillatory behavior as $Q\to\infty$
what contradicts the idea of the monotonic electric potential saturation
\cite{Tellez}.

\renewcommand{\theequation}{6.\arabic{equation}}
\setcounter{equation}{0}

\section{Conclusion}
Let us summarize briefly the crucial results of the present work.

Section 3 deals with the case of one point-like particle 
of charge $Q$ (with $\vert Q\vert < 2/\beta$) immersed 
in the bulk of the stable 2D Coulomb gas.
Passing to the sine-Gordon representation, we were able to relate
in Eq. (\ref{3.3}) the excess chemical potential $\mu_Q^{\rm ex}$
of the guest charge to the expectation value of the exponential field.
The explicit form of the latter quantity was conjectured by
Lukyanov and Zamolodchikov \cite{Lukyanov1}, see Eqs. (\ref{3.4}) and 
(\ref{3.5}), and subsequently verified by various methods.
Our Coulomb-gas formulation provides two other checks of this
conjecture: the guest-charge collapse test at $\vert Q\vert = 2/\beta$ 
and the predicted singular behavior (\ref{3.11}) of the exponential-field
expectation close to the collapse value of 
$Q=(2/\beta)-\epsilon$ $(\epsilon\to 0^+)$.  

The problem of two guest point-like particles, charged by $Q_1$
and $Q_2$ and being at distance $r$ from one another, is studied
in Section 4.
Using the form-factor method for the two-point correlation functions 
in the sine-Gordon formulation of the problem, we have derived 
the explicit formula (\ref{4.13}) for the effective interaction energy 
of the two guest charges $E_{Q_1 Q_2}(r)$ at asymptotically large distance 
$r\to\infty$.  
This formula is valid rigorously for 
$\vert Q_1\vert, \vert Q_2\vert < 2/\beta$; 
in a subspace of this region of charge values we have noticed
an anomalous weakening of the effective interaction when one
of the guest charges is increasing.
Since in the definition (\ref{4.4}) of $E_{Q_1 Q_2}(r)$ the
divergences of chemical potentials (caused by the collapse of guest
charge with electrolyte counterions) cancel with each other,
we have suggested an extended validity of the formula (\ref{4.13})
for arbitrary values of $Q_1$ and $Q_2$ (the regularization hypothesis).
Then, under certain circumstances (especially, the inequality 
$\vert Q_1\vert \ne \vert Q_2\vert$ must hold and one of the guest charges
has to be large enough), the result (\ref{4.13}) implies an effective 
attraction (repulsion) between like-charged (oppositely-charged) 
guest particles.

The adequacy of the concept of renormalized charge was confirmed in
Section 5.
This is not a surprise: in the whole stability interval of inverse
temperatures $0\le\beta<2$, the large-distance behavior of two-point
correlators is determined by the form-factor of the same particle
from the sine-Gordon spectrum, namely the $B_1$-breather with
the lightest mass $m_1$ playing the role of the inverse correlation length
of electrolyte species.
The large-distance behavior of the induced electric potential (\ref{5.5}),
considered in terms of the dimensionless combination $m_1 r$, 
is therefore basically the same in the Debye-H\"uckel limit $\beta\to 0$
as well as at every point $\beta$ which belongs to the stability interval,
up to the renormalized-charge prefactor.
The renormalized charge $Q_{\rm ren}$, considered as a function
of the (positive) bare charge $Q$, exhibits a maximum at 
$Q = (2/\beta)-(1/2)$ which is in the range $\vert Q\vert < 2/\beta$
of the rigorous validity of the formula (\ref{5.8}). 
Under the assumption of validity of the regularization hypothesis,
increasing $Q$ can produce the effect of charge inversion.
Going with $Q\to\infty$ in Eq. (\ref{5.8}) does not imply the
saturation of $Q_{\rm ren}$ at some finite value.
Instead, $Q_{\rm ren}$ oscillates between two finite extremes.
The same property holds for the electric potential in the electrolyte
region which contradicts the idea of the monotonic potential saturation
\cite{Tellez}.

The previous results obtained at the free-fermion (collapse) point $\beta=2$ 
\cite{Samaj1} are different from the present ones concerning 
the stability interval $0\le\beta<2$ in the following aspects. 
Firstly, the concept of renormalized charge fails at $\beta=2$.
Secondly, at $\beta=2$, when the bare charge $Q\to\infty$ the
induced electrostatic potential saturates {\em monotonically} at a finite
value in each point of the electrolyte region.
The reason for the fundamental differences is obvious.
The lightest $B_1$-breather disappears from the particle spectrum
of the sine-Gordon model just at $\beta=2$, and the asymptotic behavior of
two-point correlation functions at this coupling constant is
governed by the soliton-antisoliton pair.
Since $m_1\to 2M$ as $\beta\to 2$, the particle mass in the
exponential decay is a continuous function of $\beta$ at $\beta=2$.
On the other side, the inverse-power-law asymptotic behavior,
which is determined by the form-factor of the dominant particle(s)
in the sine-Gordon spectrum, undertakes an abrupt modification
when passing through the $\beta=2$ point. 
The basic qualitative features of the results obtained at 
the free-fermion point $\beta=2$ are expected to be present also for 
such $\beta>2$ where the soliton-antisoliton pair exists.
It is known \cite{Zamolodchikov1} that the soliton-antisoliton pair 
disappears from the sine-Gordon particle spectrum (and the sine-Gordon 
theory ceases to be massive) at the point $b^2=1$ $(\beta=4)$ which
corresponds to the Kosterlitz-Thouless transition of infinite order 
from the conducting (fluid) phase to the insulating phase.
We conclude that the 2D results obtained in the weak-coupling regime
of the Coulomb gas $0\le \beta<2$ differ substantially from those in 
the strong-coupling regime $2\le \beta<4$.

\renewcommand{\theequation}{A.\arabic{equation}}
\setcounter{equation}{0}

\section*{Appendix}
According to the conjectured Eqs. (\ref{3.4}) and (\ref{3.5}), it holds
\begin{equation} \label{A.1}
\lim_{\epsilon\to 0^+} \left\langle \exp\left[ {\rm i} 
\left( \frac{1}{2 b}-\epsilon b \right) \phi \right] \right\rangle
\sim \left[ \frac{\pi z \Gamma(1-b^2)}{\Gamma(b^2)} 
\right]^{\frac{1}{4 b^2 (1-b^2)}} \exp \left( I_1 + I_2 + I_3 \right)
\end{equation}
where
\begin{eqnarray}
I_1 & = & \int_0^1 \frac{{\rm d}t}{t} \left\{
\frac{{\rm sinh}(t)}{2 {\rm sinh}(b^2 t){\rm cosh}[(1-b^2)t]}
- \frac{1}{2 b^2} {\rm e}^{-2 t} \right\} \label{A.2} \\
I_2 & = & \int_1^{\infty} \frac{{\rm d}t}{t} \left\{
\frac{{\rm sinh}(t)}{2 {\rm sinh}(b^2 t){\rm cosh}[(1-b^2)t]} - 1
- \frac{1}{2 b^2} {\rm e}^{-2 t} \right\} \label{A.3} \\
I_3 & = & \int_1^{\infty} \frac{{\rm d}t}{t} \exp(-4 b^2 \epsilon t)
= -C - {\rm ln}(4 b^2 \epsilon) + O(\epsilon) \label{A.4}
\end{eqnarray}
and $C$ is the Euler's constant.
Using the same Eqs. (\ref{3.4}) and (\ref{3.5}) for expressing the expectation
value of the exponential field on the rhs of Eq. (\ref{3.11}), one gets
\begin{equation} \label{A.5}
\lim_{\epsilon\to 0^+} \frac{\left\langle \exp\left[ {\rm i} 
\left( \frac{1}{2 b}-\epsilon b \right) \phi \right] \right\rangle}{
\left\langle \exp\left[ {\rm i} 
\left( \frac{1}{2 b}- b \right) \phi \right] \right\rangle}
\sim \frac{\pi z}{2 b^2 \epsilon} \frac{\Gamma(1-b^2)}{\Gamma(b^2)}
\exp \left( I'_1 + I'_2 - C - {\rm ln} 2 \right)
\end{equation}
where
\begin{eqnarray}
I'_1 & = & \int_0^1 \frac{{\rm d}t}{t} \left\{
\frac{{\rm sinh}[(1-2 b^2)t]}{{\rm sinh}(t)} + 1
- 2 (1- b^2) {\rm e}^{-2 t} \right\} \label{A.6} \\
I'_2 & = & \int_1^{\infty} \frac{{\rm d}t}{t} \left\{
\frac{{\rm sinh}[(1-2 b^2)t]}{{\rm sinh}(t)}
- 2 (1- b^2) {\rm e}^{-2 t} \right\} \label{A.7}
\end{eqnarray}
With the aid of the integral representations \cite{Gradshteyn}
\begin{eqnarray}
C & = & \int_0^1 \frac{{\rm d}t}{t} \left( 1 - {\rm e}^{-t} \right)
- \int_1^{\infty} \frac{{\rm d}t}{t} {\rm e}^{-t} \label{A.8} \\
{\rm ln} 2 & = & \int_0^{\infty} \frac{{\rm d}t}{t} 
\left( {\rm e}^{-t} - {\rm e}^{-2 t} \right) \label{A.9}
\end{eqnarray}
the argument of the exponential on the rhs of Eq. (\ref{A.5})
can be written as
\begin{equation} \label{A.10}
I'_1 + I'_2 - C - {\rm ln} 2 =
\int_0^{\infty} \frac{{\rm d}t}{t} \left\{
\frac{{\rm sinh}[(1-2 b^2)t]}{{\rm sinh}(t)}
+ ( 2 b^2 - 1 ) {\rm e}^{-2 t} \right\} 
\end{equation}
Finally, considering in Eq. (\ref{A.5}) the integral representation of 
the logarithm of the Gamma function \cite{Gradshteyn}
\begin{equation} \label{A.11}
{\rm ln} \Gamma(x) = \int_0^{\infty} \frac{{\rm d}t}{t} {\rm e}^{- t}
\left[ ( x - 1 ) + \frac{{\rm e}^{-(x-1)t}-1}{1-{\rm e}^{-t}} \right] ,
\qquad {\rm Re}(x) > 0
\end{equation}
the proof of the desired formula (\ref{3.11}) becomes accomplished.

\section*{Acknowledgments}
I thank Bernard Jancovici for careful reading of the manuscript
and useful comments.
The support by Grant VEGA 2/3107/2003 is acknowledged.

\end{document}